\documentclass[aps,pre,twocolumn,groupedaddress,showpacs]{revtex4}

\usepackage{graphicx}

\begin{document}

\newcommand{\rbc}{Rayleigh-B\'enard convection}
\newcommand{\dotprod}{\,{\scriptscriptstyle \stackrel{\bullet}{{}}}\,}
\newcommand{\gtapprox}{\,{\scriptscriptstyle \stackrel{>}{\sim}}\,}
\newcommand{\ltapprox}{\,{\scriptscriptstyle \stackrel{<}{\sim}}\,}
\renewcommand{\vec}{\mathbf}
\newcommand{\figwidth}{3in}
\newcommand{\sech}{\mbox{sech}}

\bibliographystyle{apsrev}

\title{Mean flow and spiral defect chaos  in \rbc}

\author{K.-H.~Chiam}
\email{ChiamKH@MailAPS.ORG}
\homepage{http://www.cmp.caltech.edu/~stchaos}

\author{M.~R.~Paul}

\author{M.~C.~Cross}
\affiliation{Nonlinear and Statistical Physics, Mail Code 114-36,
  California Institute of Technology, Pasadena, CA 91125-3600}

\author{H.~S.~Greenside}
\affiliation{Department of Physics, P.~O.~Box 90305, Duke University,
  Durham, NC 27708-0305}

\date{\today}

\begin{abstract}
We describe a numerical procedure to construct a modified velocity
field that does not have any mean flow.  Using this procedure, we
present two results.  Firstly, we show that, in the absence of mean
flow, spiral defect chaos collapses to a stationary pattern comprising
textures of stripes with angular bends.  The quenched patterns are
characterized by mean wavenumbers that approach those uniquely
selected by focus-type singularities, which, in the absence of mean
flow, lie at the zig-zag instability boundary.  The quenched patterns
also have larger correlation lengths and are comprised of rolls with
less curvature.  Secondly, we describe how mean flow can contribute to
the commonly observed phenomenon of rolls terminating perpendicularly
into lateral walls.  We show that, in the absence of mean flow, rolls
begin to terminate into lateral walls at an oblique angle.  This
obliqueness increases with Rayleigh number.
\end{abstract}


\maketitle

\section{Introduction}
\label{se:intro}
\rbc\ in a horizontal layer of fluid heated from below provides a
canonical example for spatially-extended systems exhibiting pattern
formation \cite{the_review} and spatiotemporal chaos
\cite{stc,stc_nature,stc_char}.  (The latter term refers to states
that are disordered in space and that show chaotic dynamics in time.)
In this article, we present results from direct numerical simulations
of \rbc\ to investigate the importance of nonlocal flow modes,
generally referred to as the mean flow
\cite{SigZip_mf,Cross_mf,Newell_mf,Newell_mf2,Newell_mf3}, in the
formation and dynamics of patterns and spatiotemporal chaos.

In a typical \rbc\ experiment, a fluid layer is confined between two
horizontal plates, and is thermally driven far from equilibrium by
maintaining the bottom plate at a temperature that is higher than that
of the top plate.  As the temperature difference is increased, the
fluid undergoes an instability to a state in which there is motion
driven by the buoyancy forces. When the temperature difference between
the plates is above but near this convective threshold, a pattern
comprised of patches of locally parallel convection rolls forms with
roll diameters that are close to the depth of the cell; see, for
example, Fig.~\ref{fig:stripetex}(a).  When the temperature difference
is increased, the fluid undergoes other instabilities that may result
in the pattern developing a simple or chaotic time-dependence.
Finally, when the temperature difference is increased further,
spatiotemporal chaotic states may appear.  In particular, a state
called spiral defect chaos \cite{MBCA} is observed for Rayleigh number
$R \gtapprox 3000$, Prandtl number $\sigma \sim 1$, and aspect ratio
$\Gamma \gtapprox 16$.  This state is a disordered collection of
spirals that rotate in both directions and coexist with dynamical
defects such as grain boundaries and dislocations (see, for example,
Fig.~\ref{fig:sdc}(a)).

The following facts are known about spiral defect chaos: The parameter
values for which it appears \cite{MBCA,Hu,Hu2,Liu,MBCA2}; the
distributions of local roll properties such as the wavenumber
\cite{Egolf}; statistics of spiral and defect populations
\cite{Egolf,Ecke_Science}; the mechanisms for the generation of chaos
from spatial disorder \cite{Egolf_Nature}; the wavenumber selection
mechanism for spirals \cite{yuhai,yuhai2}; and the conditions under
which spiral defect chaos transitions to other states.  Of particular
interest to this article are experiments \cite{Assen1,Assen2} that
have observed that the spirals transform into targets when the Prandtl
number is increased from $\sigma \sim 1$ to $\sigma \sim 10$ and when
the Rayleigh number $R \gtapprox 3500$.  While this observation
establishes that spiral defect chaos occurs only at low Prandtl
numbers, it does not allow us to conclude which of the many dynamic
phenomena that occur at low Prandtl numbers \cite{the_review} is
responsible for the formation of spiral defect chaos.

One particular phenomenon that becomes important at low Prandtl
numbers is the presence of mean flow.  Mean flow is the name given to
the velocity field with a non-zero mean over the depth of the
convective cell that is generated by the variations of the structure
of the convection rolls such as their curvature, amplitude, and
wavenumber, and that in turn couples through advection to further
modify the roll structure.  Its magnitude is approximately inversely
proportional to the Prandtl number \cite{CN}.

It is believed that spiral defect chaos is an effect of mean flow which in turn
is important at low Prandtl numbers \cite{MBCA,MBCA2}. This hypothesis has been
investigated in numerical studies of model equations of convection.  For
example, by coupling a mean flow-like field to the Swift-Hohenberg equation
\cite{SH}, chaotic behavior is observed \cite{HSG_mf}.  Furthermore, when the
parameter in the model that gives the strength of the mean flow is made large,
spatiotemporal chaotic states akin to spiral defect chaos is observed
\cite{Xi,Xi2,Xi3,Tsimring}.  However, the above results are tenuous because of
two reasons.  Firstly, the Swift-Hohenberg equation describes the
spatiotemporal behavior of a field in two dimensions, whereas convection is a
three-dimensional phenomenon.  Secondly, there are known limitations to the
Swift-Hohenberg modelling.  For example, it is known that the long-time
dynamics of the Swift-Hohenberg equation may not correspond to that of the
Boussinesq equations of convection.  In particular, Swift-Hohenberg models
exhibit spiral defect chaos as a transient behavior, whereas in experiments,
spiral defect chaos is known to persist for much longer times \cite{SH_limit}.
In addition, it is also known that the small-scale structure of the mean flow
at the cores of the spirals, which might be crucial for the persistence of
spiral defect chaos, is not perfectly captured in the Swift-Hohenberg equation
\cite{SH_limit}.

Our goal in this article is to show by direct numerical simulations of
\rbc\ that spiral defect chaos is indeed a consequence of the presence
of mean flow.  In the absence of mean flow, we find that spiral defect
chaotic states cease to exist, and are replaced by states whose
statistical properties differ from those of spiral defect chaos.  In
general, studies of mean flow are difficult to perform in experiments,
primarily because it is difficult to measure mean flow in an
experimental setup.  This is due to several reasons, namely that the
magnitude of mean flow is small (typically on the order of $1\%$ of
the magnitude of the velocity of the convecting rolls), and that it
exists only in distorted and not regular patterns.  To the best of our
knowledge, there has only been one experiment that has successfully
imaged aspects of the mean flow, but only in a simple distorted
pattern \cite{french_paper,french_paper2}.  It is not clear if such
imaging techniques can be applied to more general and complicated
patterns.  Thus, direct numerical simulations are particularly
valuable for the study of mean flow.

We achieve this goal by numerically constructing a \emph{gedanken}
fluid whose velocity field is modified to have zero mean flow.  By
investigating the states that arise from the dynamics of this fluid
and by comparing them with spiral defect chaos, we can infer directly
the role of mean flow in the formation and dynamics of spiral defect
chaos.

Once we have the capability to remove mean flow from the fluid
dynamics, we can apply it to the study of other problems.  One such
problem that we have investigated is the relation between mean flow
and lateral boundaries.  Using our numerical simulations, we have
shown for the first time how mean flow can contribute to the commonly
observed phenomenon of convection rolls terminating perpendicularly
into lateral walls, an observation that is still without much
theoretical understanding.

The remaining of this article is organized as follows: In
Section~\ref{se:def}, we define the equations governing \rbc, what a
mean flow is, and how it can be measured and eliminated numerically.
In Section~\ref{se:res}, we present results on the relation between
mean flow and spiral defect chaos, stripe textures, and lateral
boundaries.  In Section~\ref{se:con}, we present our conclusions.

\section{Definitions}
\label{se:def}
\subsection{Boussinesq Equations}

The evolution of a low-velocity and hence approximately incompressible
convecting fluid is governed to good approximation by the three-dimensional
Boussinesq equations \cite{the_review}.  They are the combination of the
incompressible Navier-Stokes and heat equations, with the further assumption
that density variations are proportional to temperature variations and that
this density variation appears only in the buoyancy force.  Written in a
dimensionless form, they are:
\begin{eqnarray}
\label{eq:NS}
\sigma^{-1} \left( \partial _{t}+\vec{u}\dotprod \vec{\nabla}
\right) \vec{u}(x,y,z,t) &=&-\vec{\nabla }p+ \vec{\nabla }^{2}\vec{u}+
RT\widehat{z}, \\
\label{eq:T}
\left( \partial _{t}+\vec{u}\dotprod \vec{\nabla} \right) T(x,y,z,t)
&=&\vec{\nabla }^{2}T, \\
\label{eq:incom} \vec{\nabla} \dotprod \vec{u} &=&0.
\end{eqnarray}
The field $\vec{u}(x,y,z,t)$ is the velocity field at point $(x,y,z)$
at time $t$, while $p$ and $T$ are the pressure and temperature fields
respectively.  The variables $x$ and $y$ denote the horizontal
coordinates, while the variable $z$ denotes the vertical coordinate,
with the unit vector $\widehat{z}$ pointing in the direction opposite
to the gravitational acceleration. The spatial units are measured in
units of the cell depth $d$, and time is measured in units of the
vertical diffusion time $d^2/\kappa$, where $\kappa$ is the thermal
diffusivity of the fluid.  The parameter $R$ is the Rayleigh number,
defined to be the dimensionless temperature difference $\Delta T$
across the top and bottom plates,
\begin{equation}
  \label{eq:R}
  R = \frac{\alpha g d^3}{\nu \kappa} \Delta T,
\end{equation}
where $\alpha$ is the thermal expansion coefficient, $\kappa$ the thermal
diffusivity, and $\nu$ the viscous diffusivity (kinematic viscosity) of the
fluid. In this article, we will also frequently use the reduced Rayleigh
number,
\begin{equation}\label{eq:reducedR}
  \epsilon = \frac{R-R_c}{R_c},
\end{equation}
where $R_c \approx 1708$ is the critical Rayleigh number at the onset
of convection in an infinite domain \cite{the_review}.  The parameter
$\sigma$ is the Prandtl number, defined to be the ratio of the fluid's
thermal to viscous diffusivities,
\begin{equation}
  \label{eq:sigma}
  \sigma = \frac{\nu}{\kappa}.
\end{equation}

The material walls are no-slip so that the velocity field satisfies
\begin{equation}
    \label{eq:u_bcs}
  \vec{u} = 0, \ \ \ \mbox{on all material walls}.
\end{equation}
The temperature field is constant on the top and bottom plates:
\begin{equation}
    \label{eq:T_tb_bcs}
  T\left(x,y,z=\mp\frac{1}{2},t\right)=\pm \frac{1}{2},
\end{equation}
and we assume that the lateral walls are perfectly insulating, so that
\begin{equation}
  \label{eq:T_lateral_bcs}
  \widehat{n}\dotprod \vec{\nabla} T = 0, \ \ \ \mbox{on lateral walls,}
\end{equation}
where $\widehat{n}$ is the unit vector perpendicular to the lateral
walls at a given point.  The pressure field $p$ has no associated
boundary condition since it does not satisfy a dynamical equation.

The influence of the lateral walls on the dynamics is determined by
the dimensionless aspect ratio $\Gamma$, defined to be the
half-width-to-depth ratio of the cell if it is rectangular and the
radius-to-depth ratio if it is cylindrical.

\subsection{Direct Numerical Simulations}
\label{se:dns}

We used two different numerical schemes to solve the Boussinesq
equations.  The first is a serial second-order-accurate
finite-difference scheme that is based on a cubic colocated mesh.  It
is highly efficient for simulating a rectangular cell of moderate
aspect ratio.  The second is a parallel spectral element scheme that
is second-order-accurate in time and is able to treat more complex
geometries with arbitrary lateral boundaries.  Both schemes were used
to obtain the results presented in this article and were found to give
good agreement with each other.  Details of both of these schemes are
available elsewhere \cite{mclai,nekton}.  For applications of these
schemes to related problems in \rbc, see
Refs.~\cite{mark1,mark2,mark3}.

\subsection{Mean Flow}
\label{se:meanflow}

When the convection pattern is made up of rolls that are not straight
and parallel, a mean flow, slowly varying in the horizontal
coordinates, will be set up.  The importance of mean flow is that it
is a nonlocal flow mode, and as such, affects the global behavior of
the convection pattern even though its magnitude is small.  A detailed
derivation of mean flow can be found in Ref.~\cite{CN}.
Heuristically, it can be understood as follows.  When there are
inhomogeneities in the amplitude $A(x,y)$ and wave vector
$\vec{k}(x,y)$ (or equivalently, the phase $\phi(x,y)$ where
$\vec{\nabla}_\perp \phi = \vec{k}$) of the convection rolls, a
Reynolds stress will be generated locally from the gradients of
$\vec{k}$ and $A$.  This results in a flow slowly varying in the
plane.  In addition, these inhomogeneities will also induce a varying
component $p_s(x,y,t)$ in the pressure field that is constant across
the depth of the cell and slowly varying in the plane.  The gradient
$\vec{\nabla}_\perp p_s$ will then drive a global flow that, together
with the Reynolds-stress-induced flow, distorts the convection rolls
further.  If we call the slowly varying flow $\vec{u}_D$, then we can
write \cite{CN}
\begin{equation}
  \label{eq:u_D}
  \sigma \partial_{zz} \vec{u}_D = \vec{\nabla}_\perp p_s +
  \frac{1}{2\pi} \int_0^{2\pi} d\phi\, \vec{u} \dotprod \vec{\nabla}
  \vec{u}_\perp,
\end{equation}
where the integral over the phase variable $\phi$ serves to average
out the fast modes of the integrand.  The as yet unknown field $p_s$
can be determined via the incompressibility condition,
Eq.~(\ref{eq:incom}), which requires that
\begin{equation}
  \label{eq:u_D_incom}
  \vec{\nabla}_\perp \dotprod \int_{-1/2}^{1/2} dz\, \vec{u}_D(x,y,z,t) = 0.
\end{equation}
Eq.~(\ref{eq:u_D}) can then be integrated twice with respect to $z$,
with boundary condition Eq.~(\ref{eq:u_bcs}), to completely give
$\vec{u}_D$.  Finally, the slow distortions, $\vec{u}_D$, advect the
phase contours of the convection rolls, yielding an additional
advection term in the phase equation \cite{CN},
\begin{equation}
  \partial_t \phi \rightarrow \partial_t \phi + \vec{U}\dotprod
  \vec{\nabla}_\perp \phi.
\end{equation}
The velocity field $\vec{U}$ is called the mean flow.  It is an
average of the slow distortions over the depth of the cell,
\begin{equation}
  \vec{U}(x,y,t) = \int_{-1/2}^{1/2}dz\, \vec{u}_D(x,y,z,t) g(z),
\end{equation}
with $g(z)$ a function that depends on the nonlinear structure of the
rolls.

We can approximate the mean flow from our numerical simulations as the
average over the depth of the cell of the slow components of the
horizontal velocity,
\begin{equation}
  \label{eq:U}
  \vec{U}(x,y,t) \approx \frac{1}{2\pi} \int_0^{2\pi} d\phi\,
  \int_{-1/2}^{1/2} dz\,  \vec{u}_\perp(x,y,z,t).
\end{equation}
In practice, we replace the integral over the phase variable $\phi$
with a Gaussian filter of characteristic width $\mathcal{O}(1)$ so
that variations over short length scales are smoothed out.

For the approximation of Eq.~(\ref{eq:U}), and with the no-slip
boundaries Eq.~(\ref{eq:u_bcs}), the mean flow $\vec{U}(x,y,t)$ is
solenoidal:
\begin{equation}
  \label{eq:U_incom}
  \vec{\nabla}_\perp \dotprod \vec{U} = 0.
\end{equation}
We will also find it convenient to use  the mean flow stream
function, $\zeta(x,y,t)$, and the vertical component of the mean flow
vorticity, $\omega_z$, defined by
\begin{equation}
  \label{eq:zeta}
  - \vec{\nabla}_\perp^2 \zeta = \omega_z = \widehat{z} \dotprod
  \left( \vec{\nabla}_\perp \times \vec{U}\right).
\end{equation}
The stream function, in particular, is useful to visualize because it
gives the stream lines and so the geometry of the mean flow.

\subsection{Quenching Mean Flow}
\label{se:quench}

We now describe a procedure to construct a modified velocity field
that does not have any mean flow.  To do this, we want to add to the
right-hand-side of Eq.~(\ref{eq:u_D}) the negative of the source of
the slow distortions, i.e., the depth average of the Reynolds stress,
so that $\vec{u}_D$ becomes zero for all $(x,y,z,t)$.  In
Appendix~\ref{se:fudge}, we show that this additional term takes the
form
\begin{equation}
  \label{eq:Phi}
  \vec{\Phi}(x,y,t) = - \rho \int_{-1/2}^{1/2} dz\,
  \vec{u} \dotprod \vec{\nabla} \vec{u}_\perp
\end{equation}
with $\rho \approx 1.5$ a constant.  We can then add $\vec{\Phi}$ to
the fluid equation, so that Eq.~(\ref{eq:NS}) becomes
\begin{equation}
  \label{eq:NS_imaginary}
\sigma^{-1} \left( \partial _{t} +\vec{u} \dotprod \vec{\nabla}
\right) \vec{u}(x,y,z,t) = -\vec{\nabla}p+ \vec{\nabla}^{2}\vec{u}+
RT\widehat{z} + \sigma^{-1}\vec{\Phi}.
\end{equation}
If $\vec{\Phi}$ is introduced at time $t=t_q$, the time needed for the
modified velocity field $\vec{u}$ to respond to this additional
forcing can be estimated by applying dimensional arguments on the
terms in Eq.~(\ref{eq:NS_imaginary}).  This time scale is
$\mathcal{O}(\sigma)$.  In this paper, we consider $\sigma = 1$ so we
expect the mean flow to be quenched in a time scale of
$\mathcal{O}(1)$ from time $t_q$.

\begin{figure}[htb]
  \begin{center}
    \includegraphics[width=\figwidth]{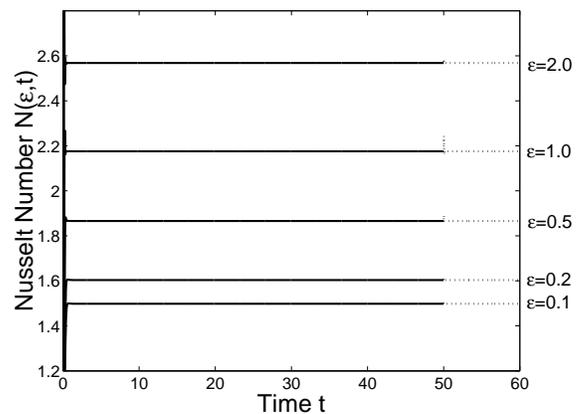}
  \end{center}
  \caption{Time series of the Nusselt number $N(\epsilon,t)$ for
    patterns comprising straight parallel rolls that have no mean flow
    at several values of $\epsilon$ before (denoted by solid lines)
    and after (dotted lines) quenching of the mean flow, which
    occurred at time $t_q=50$.  All data reported here are for Prandtl
    number $\sigma=1$ and a rectangular cell of aspect ratio
    $\Gamma_x=\Gamma_y=20$.  (In order to achieve straight parallel
    rolls, periodic lateral boundaries were imposed and a small
    sinusoidal perturbation in the temperature field was used as the
    initial condition.)}
  \label{fig:nusselt_isr}
\end{figure}

For a pattern that does not have mean flow, such as a pattern
comprising straight parallel rolls with no defects or concentric
circular rolls, the quenching procedure should leave the convective
properties, such as the Nusselt number, of the fluid unchanged.  The
Nusselt number is the ratio of convective heat transfer to heat
transfer that would occur by conduction alone if the fluid remained at
rest.  In Fig.~\ref{fig:nusselt_isr}, we show that this is true.  The
Nusselt number before (denoted by solid lines) and after (dashed
lines) the quenching procedure, which occurred at $t=t_q=50$, are
indeed the same.

\section{Results}
\label{se:res}
\subsection{Mean Flow and Spiral Defect Chaos}

\begin{figure}[htb]
  \begin{center}
    \includegraphics[width=\figwidth]{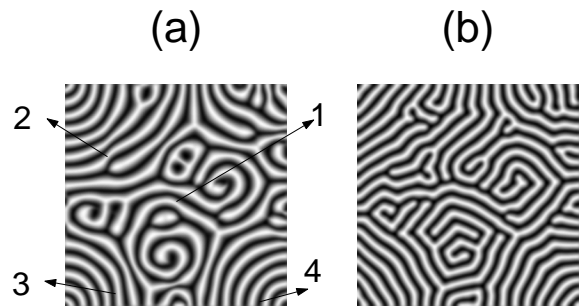}
  \end{center}
  \caption{\emph{(a)} An example of spiral defect chaos observed in a
    numerical simulation using the finite-difference scheme
    \cite{mclai}.  The mid-plane temperature field is plotted at time
    $t=500$ for parameters $\epsilon=1.0$, $\sigma=1$, and
    $\Gamma_x=\Gamma_y=20$.  Dark regions correspond to cold sinking
    fluid, light regions to hot rising fluid.  The spiral defect chaos
    planform is characterized by a disordered collection of spirals
    rotating in both directions and coexisting with dynamical defects
    such as grain boundaries and dislocations. The labels ``1'' to
    ``4'' are discussed in Fig.~\ref{fig:sdc_time_series}.  \emph{(b)}
    When mean flow is quenched, spiral defect chaos collapses to a
    stationary pattern of textures of stripes with angular bends.  The
    planform shown here is at $10$ time units after the quenching has
    been introduced to the state shown in (a).  All other parameters
    are unchanged.}
  \label{fig:sdc}
\end{figure}

Using the numerical schemes described in Section~\ref{se:dns}, we
evolved
Eqs.~(\ref{eq:T}),~(\ref{eq:incom}),~and~(\ref{eq:NS_imaginary}) from
the initial conditions
\begin{equation}
\label{eq:random_init}
\vec{u}(x,y,z,t=0)=p(x,y,z,t=0)=0,
\end{equation}
and
\begin{equation}
\label{eq:random_init2}
T(x,y,z,t=0)=-z + \eta(x,y,z),
\end{equation}
where $T=-z$ is the linear conduction profile and $\eta$ is randomly
chosen from a uniform distribution in the range $[-10^{-5},10^{-5}]$.
We observed spiral defect chaos when the parameters are chosen such
that the reduced Rayleigh number $\epsilon$ lies in the range
$[0.6,3.0]$, the Prandtl number $\sigma \approx 1$, and the aspect
ratio lies in the range $[16,30]$.  In Fig.~\ref{fig:sdc}(a), we show
an example: a planform of the mid-plane temperature field $T(x,y,z=0)$
at time $t=500$ for parameters $\epsilon=1.0$, $\sigma=1$, and
$\Gamma_x=\Gamma_y=20$.  In general, the planforms we observed are
qualitatively similar to those observed in experiments in both
cylindrical \cite{MBCA} and rectangular \cite{rect_SDC} geometries.

We note that the range of aspect ratios that we have simulated are
smaller that those of past experiments which extend up to $\Gamma
\gtapprox 50$ \cite{MBCA,Hu2,MBCA2}.  This limitation is caused by the
need to integrate up to at least the horizontal diffusion time scale,
$t \sim \Gamma^2$, which is an estimate of the minimum time for
thermal transients to diffuse over the entire cell and thus for the
pattern to reach an asymptotic state.  Because of this quadratic
dependence on the aspect ratio, the time needed to execute the
numerical schemes so that an asymptotic pattern is reached becomes
prohibitively long for $\Gamma$ beyond about $20$~to~$30$ (based on
current computational resources; for details, see Ref. \cite{mclai}).

\begin{figure}[htb]
  \begin{center}
    \includegraphics[width=\figwidth]{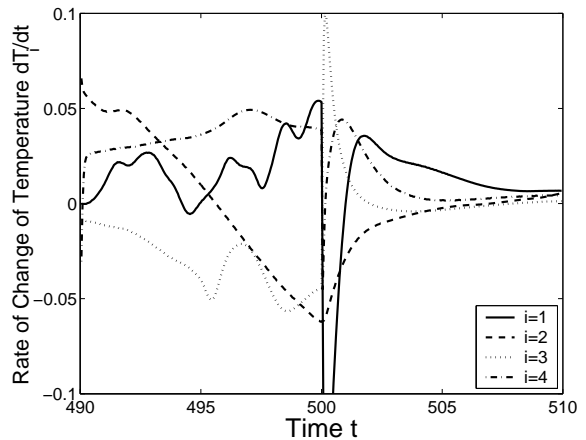}
  \end{center}
  \caption{The rate of change of the temperature field $dT_i/dt$
    versus time $t$ for the four locations in the cell indicated in
    Fig.~\ref{fig:sdc}(a).  Prior to quenching of the mean flow which
    takes place at time $t=t_q=500$, the derivative $dT/dt$ fluctuates
    and differs from zero.  After quenching, it approaches zero in a
    time scale of $\mathcal{O}(1)$, suggesting that the pattern is
    approaching stationarity.}
  \label{fig:sdc_time_series}
\end{figure}

In the rest of this Section, we report on results simulated in a
rectangular cell of aspect ratio $\Gamma_x=\Gamma_y=20$.  We
integrated for $500$ time units, and then, at time $t=t_q=500$,
invoked the forcing term given by Eq.~(\ref{eq:Phi}) that will quench
the mean flow dynamics.  In Fig.~\ref{fig:sdc}(b), we plot the
mid-plane temperature field at time $t=510$ which is ten time units
after the quenching of the mean flow has begun.  (Recall that the
quenching takes place in a time of $\mathcal{O}(\sigma)$ so the
quenched state at ten time units should have already been asymptotic
for our $\sigma=1$ state here.)  We see that the rolls have
``straightened out'' in that they have lost their curvature and have
developed angular bends.  More strikingly, the straightened roll
patches become stationary, leaving the only dynamics in the pattern to
come from the motion of defects such as dislocations and grain
boundaries.  To illustrate this transition from a dynamical state to a
seemingly ``frozen'' one, we plot in Fig.~\ref{fig:sdc_time_series}
the time series of the rate of change of the temperature field at
several locations in the cell.  We see that, for $t<t_q=500$, the
derivative $dT(x,y)/dt$ fluctuates and is significantly different than
zero at all $t<t_q$.  However, after the quenching of the mean flow is
initiated at $t=t_q=500$, the derivative $dT(x,y)/dt$ relaxes to
approach zero in a time scale of $\mathcal{O}(1)$, suggesting that all
dynamics is becoming ``frozen'' and that a stationary pattern is being
approached.

\begin{figure}[htb]
  \begin{center}
    \includegraphics[width=\figwidth]{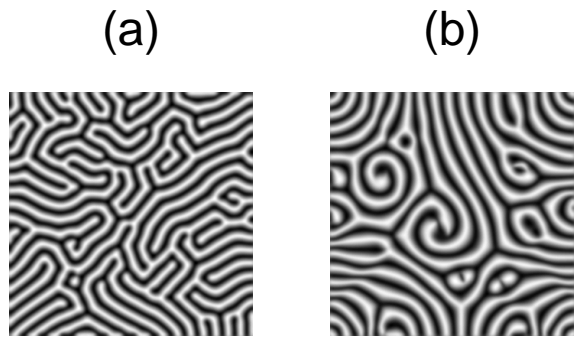}
  \end{center}
  \caption{\emph{(a)} Stationary patches of  stripes with angular bends
    at time $t=100$ when mean flow quenching is introduced at time
    $t=0$.  The parameters are $\epsilon=1.0$, $\sigma=1$, and
    $\Gamma_x=\Gamma_y=20$. \emph{(b)} When the quenching is turned
    off at time $t=100$ so that mean flow is restored, spiral defect
    chaos is observed.  The planform shown here is at $500$ time units
    after the restoration of mean flow.}
  \label{fig:sdc_random_init}
\end{figure}

We have also repeated the quenching of the mean flow at other Rayleigh
numbers ranging from $\epsilon=0.6$~to~$3.0$, and for different
instances of the initial condition Eq.~(\ref{eq:random_init2}).  In
all cases, we observed similar stationary planforms as shown in
Fig.~\ref{fig:sdc}(b).  In addition, this spiral-to-angular transition
can be observed in the reverse direction.  When the mean flow
quenching is turned off at a later time $t=550$ so that mean flow is
again restored to the system, the angular bends develop into spirals
and the stationary planform becomes dynamical again.  Spiral defect
chaos is fully restored \cite{movie}.  Furthermore, the stationary
textures of stripes with angular bends can also be observed when the
quenching is initiated at other times.  For example, instead of
initiating the mean flow quenching procedure at a time when a spiral
defect chaotic state is already asymptotic, we have also initiated the
quenching procedure immediately at the start of the simulation,
$t=t_q=0$, again using
Eqs.~(\ref{eq:random_init})-(\ref{eq:random_init2}) as initial
conditions.  In Fig.~\ref{fig:sdc_random_init}(a), we show the
planform after $100$ time units for the parameters $\epsilon=1.0$,
$\sigma=1$, and $\Gamma_x=\Gamma_y=20$.  We see that it comprises
patches of locally straight rolls ending into each other in angular
bends.  There are no spirals present.  When mean flow is restored at
time $t=100$, we find that, after a time of $\mathcal{O}(1)$, spiral
defect chaos appears, as can be seen in
Fig.~\ref{fig:sdc_random_init}(b) which shows the planform at $500$
time units after the mean flow has been restored.

Thus, we have shown that spiral defect chaos does not exist without the
presence of mean flow.

\begin{figure}[htb]
  \begin{center}
    \includegraphics[width=\figwidth]{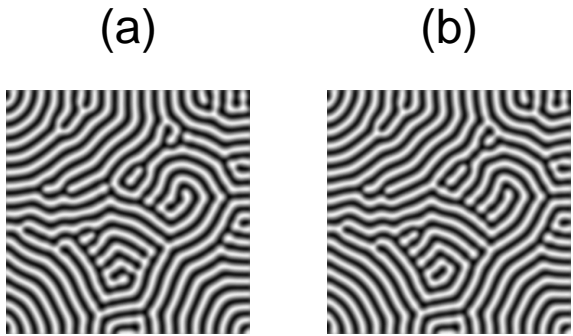}
  \end{center}
  \caption{\emph{(a)} The pattern observed when the Prandtl number is
    instantaneously increased from $\sigma=1$ to $\sigma=10$ comprises
    stripes with angular bends that are similar to the quenched
    patterns in Fig.~\ref{fig:sdc}(a).  The pattern shown here is at
    $100$ time units after the Prandtl number has been instantaneously
    increased.  The parameters correspond to those of the state in
    Fig.~\ref{fig:sdc}.  \emph{(b)} When the mean flow is quenched for
    the $\sigma=10$ state of (a), we see that the resulting pattern is
    qualitatively unchanged.  Shown here is the state at $100$ time
    units after the mean flow has been quenched.}
  \label{fig:sdc_highpr}
\end{figure}

Before we conclude this section, we qualitatively compare the
differences between the states observed when mean flow is quenched and
at high Prandtl numbers, for which mean flow is weak.  (Recall that
the magnitude of mean flow is inversely proportional to the Prandtl
number.)  Starting from the state shown in Fig.~\ref{fig:sdc}(a), we
instantaneously increased the Prandtl number for that state from
$\sigma=1$ to $\sigma=10$ at time $t=500$.  Although increasing the
Prandtl number changes the convective properties of the fluid and
hence the dynamics of the state, we nevertheless observed (see
Fig.~\ref{fig:sdc_highpr}(a)) stripes with angular bends that are
similar to those observed when the mean flow is quenched.  Thus, the
states observed when the mean flow is quenched and unquenched states
observed at high Prandtl numbers are similar.  In addition, we also
show in Fig.~\ref{fig:sdc_highpr}(b) the state observed when we invoke
the mean flow quenching procedure after increasing the Prandtl number
to $\sigma=10$.  We see that it is again similar to the pattern at
$\sigma=10$, suggesting that even at $\sigma=10$, the residual mean
flow components are negligible.

Finally, we note that, contrary to the results of Assenheimer and
Steinberg \cite{Assen1,Assen2}, we do not observe the transition from
spirals to targets as the Prandtl number is increased to $\sigma =10$.
Several explanations are plausible: Firstly, in the Assenheimer and
Steinberg experiments, non-Boussinesq effects are significant
\cite{Assen1,Assen2}, whereas our direct numerical simulations are
only for Boussinesq fluids.  Secondly, our smaller aspect ratios may
not support the formation of targets, and that we would indeed see the
spiral to target transition in larger aspect ratios.  Thirdly, the
transition to targets may be strongly dependent on the history of the
system, in particular on the path (in system space) that the
parameters traverse.

\subsection{Mean Flow and Pre-chaotic Stripe Textures}

\begin{figure}[htb]
  \begin{center}
    \includegraphics[width=\figwidth]{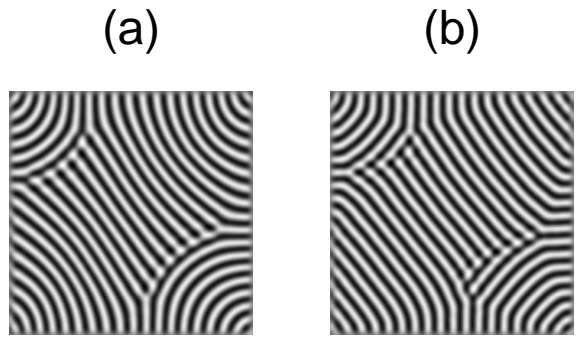}
  \end{center}
  \caption{\emph{(a)} Mid-plane temperature field at time $t=500$ for
    parameters $\epsilon=0.15$, $\sigma=1$ and $\Gamma_x=\Gamma_y=20$.
    The stripe texture comprises patches of locally parallel rolls and
    arcs that are stationary.  \emph{(b)} The state of (a) observed at
    $10$ time units after the mean flow is quenched.  The curved rolls
    have transitioned into stripes with angular bends that are
    stationary.}
  \label{fig:stripetex}
\end{figure}

At lower Rayleigh numbers near the convective threshold, the planforms
observed take the form of stripe textures rather than exhibiting
spiral defect chaos.  They comprise patches of locally parallel rolls
and arcs such that each patch terminates at the boundaries of another
at a different orientation, and the boundaries between the patches are
usually populated by defects.  In general, the stripe textures are
stationary after transients, except for the motion of defects at the
grain boundaries.  In Fig.~\ref{fig:stripetex}(a), we show a planform
of the mid-plane temperature field at time $t=500$ at $\epsilon=0.15$
and $\sigma=1$ in a rectangular cell of aspect ratio
$\Gamma_x=\Gamma_y=20$ .

When the mean flow is quenched at time $t=t_q=500$, we observe that
the stationary stripe textures remain stationary, and that those rolls
that are curved are straightened out.  The resulting pattern, shown in
Fig.~\ref{fig:stripetex}(b) which is at $10$ time units after the
quenching, comprises patches of angular structures that replaced
patches of curved arcs.

\subsection{Nusselt Numbers}

\begin{figure}[htb]
  \begin{center}
    \includegraphics[width=\figwidth]{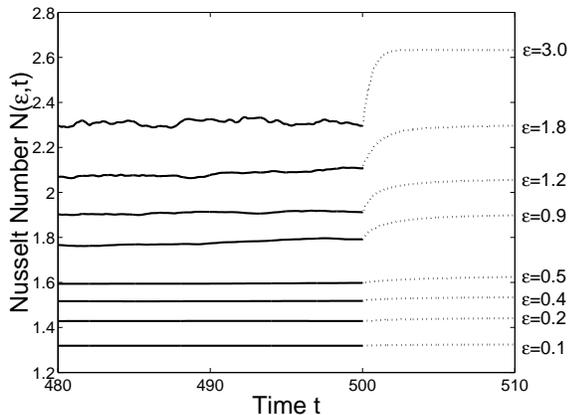}
  \end{center}
  \caption{Time series of the Nusselt number $N(\epsilon,t)$ for
    stripe textures and spiral defect chaos at several values of
    $\epsilon$ before (denoted by solid lines) and after (dotted
    lines) quenching of the mean flow which occurred at time
    $t_q=500$.  All data reported here are for Prandtl number
    $\sigma=1$ and a rectangular cell of aspect ratio
    $\Gamma_x=\Gamma_y=20$.}
  \label{fig:nusselt}
\end{figure}

One way to quantify the changes introduced by the quenching procedure
to a pattern is to look at its global convective properties, such as
the Nusselt number.  For a pattern with mean flow, its Nusselt number
will be different than those of the unmodified velocity field of
Eq.~(\ref{eq:NS_imaginary}) because the latter is not a solution to
the Boussinesq equations.  An alternate way of saying this is that
Eq.~(\ref{eq:NS_imaginary}), together with
Eqs.~(\ref{eq:T})~and~(\ref{eq:incom}), can be interpreted as the
driven Boussinesq equations with a driving force
$\sigma^{-1}\vec{\Phi}$ that is turned on at time $t_q$.  Owing to
this driving, we expect the convective properties of the fluid to be
stronger at time $t>t_q$ than at time $t<t_q$.  This is illustrated in
Fig.~\ref{fig:nusselt}.  The fractional change in the Nusselt number
$\Delta N/N$ caused by the introduction of the quenching of the mean
flow increases with the reduced Rayleigh number.  A best linear fit to
the data yields the relation
\begin{equation}
  \Delta N/N = (0.052 \pm 0.005) \epsilon.
\end{equation}
Thus, for example, when $\epsilon \sim 1$, modifying the velocity
field to quench the mean flow introduces a change of approximately
$5\%$ to the averaged convective properties of the fluid.

\subsection{Wavenumber Distributions}

In this section, we quantify the differences between the patterns
observed with mean flow and with mean flow quenched by studying the
wavenumber distributions.  We compute the probability density function
of wavenumbers, $P(k)$, from a time-average of the patterns.  We used
the local method discussed in Ref.~\cite{Egolf} to calculate the
wavenumber distributions.  We have found that, for smaller aspect
ratios $\Gamma \ltapprox 20$, this method produces better statistics
than global Fourier transform methods that were used in previous
experiments \cite{MBCA,MBCA2,Hu2}.  The mean of the wavenumber
distribution then gives the mean wavenumber $\langle k \rangle
(\epsilon)$ as a function of the reduced Rayleigh number $\epsilon$.

\begin{figure}[htb]
  \begin{center}
    \includegraphics[width=\figwidth]{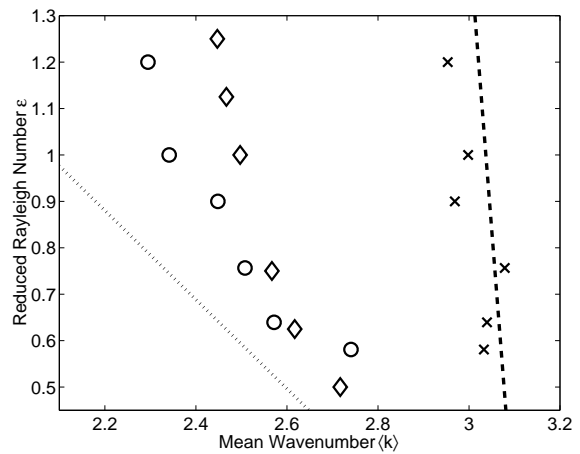}
  \end{center}
  \caption{Mean  wavenumbers $\langle k \rangle$
    for various reduced Rayleigh numbers $\epsilon$.  The circle
    symbols denote wavenumbers estimated for spiral defect chaotic
    states at $\sigma=1$ and $\Gamma_x=\Gamma_y=20$ from our direct
    numerical simulations averaged over different random initial
    conditions.  The cross symbols denote wavenumbers for states
    observed at $10$ time units after mean flow is quenched.  For
    comparison purposes, the diamond symbols denote wavenumbers
    obtained in the experiment of Ref.~\cite{MBCA} in a cylindrical
    cell of $\Gamma=78, \sigma=0.95$.  The dashed line denotes the
    unique wavenumber $k_f$ possessed by focus-selected convection at
    $\sigma=1$, and is represented by Eq.~(\ref{eq:k_f}).  The dotted
    line denotes the unique wavenumber $k_d$ selected by
    dislocations.}
  \label{fig:wavenumbers}
\end{figure}

Before we highlight the differences, we point our that the mean
wavenumbers obtained from our numerical simulations of spiral defect
chaos lie within the Busse stability balloon
\cite{the_review,Busse}. In addition, they are also consistent with
existing theory for the selection of wavenumbers in spiral defect
chaos \cite{yuhai,yuhai2}, which suggests that the wavenumbers of
convecting spirals are ``frustrated'', i.e., they lie between two
competing selection mechanisms, selection by focus-type singularities
\cite{BC} and selection by dislocations \cite{q_d,annu}.  These two
sets of selected wavenumbers, at $\sigma=1$, are denoted in
Fig.~\ref{fig:wavenumbers} by the dashed and the dotted lines,
respectively.  We see that our direct numerical simulations produced
wavenumbers (denoted by the circle symbols) that lie within these two
sets of selected wavenumbers.  For comparison purposes, we have also
included the mean wavenumbers calculated in a previous experiment
\cite{MBCA,MBCA2} performed in a cylindrical cell with $\Gamma=78$ and
$\sigma=0.95$ (diamond symbols).  We see that, at lower Rayleigh
numbers, the mean wavenumbers from our simulations agree with the
experimental findings.  However, at higher Rayleigh numbers, the
wavenumbers from our simulations are smaller than those of the
experiments.  Presumably, the smaller aspect ratios used in our
simulations mean that our wavenumbers are affected by finite size
effects.

For the range $0.6 \le \epsilon \le 1.2$, the mean wavenumbers of the
stripes with angular bends when mean flow is quenched (denoted by the
cross symbols in Fig.~\ref{fig:wavenumbers}) appear to fall onto a
straight line whose mathematical form can be obtained from a linear
fit,
\begin{equation}
  \label{eq:k_quenched}
  \langle k \rangle = (3.14 \pm 0.05) - (0.16 \pm 0.06) \ \epsilon.
\end{equation}
This relation is consistent with the wavenumbers selected by
focus-type singularities at Prandtl number $\sigma=1$ \cite{BC},
\begin{equation}
  \label{eq:k_f}
  k_f = 3.117 -0.13 \ \epsilon.
\end{equation}
The local pattern in focus-selected convection includes rolls that
form closed contours about a point within the cell.  In our
rectangular geometries, the four corners act as focus centers, as can
be seen by the presence of approximately axisymmetric roll patches
emanating from the corners, see Fig.~\ref{fig:sdc}(a).  In the absence
of mean flow, the wavenumber selected therefore appears to be
dominated by that selected by the focus centers (i.e., the corners) to
give a mean wavenumber consistent with that selected in focus-type
singularities.

Furthermore, in the absence of mean flow, the wavenumbers $k_f$ lie at
the boundary of the zig-zag instability \cite{CN}.  The patterns
observed with mean flow quenched are thus dominated by lateral ``zig
and zag'' bendings, leading to the stripes with angular bends observed
in Figs.~\ref{fig:sdc}(b)~and~\ref{fig:stripetex}(b).

\begin{figure}[htb]
  \begin{center}
    \includegraphics[width=\figwidth]{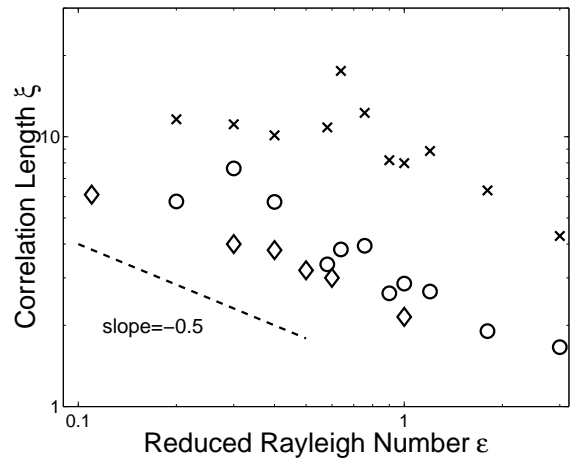}
  \end{center}
  \caption{Correlation length $\xi$ vs. reduced Rayleigh number
    $\epsilon$.  The symbols are as defined in
    Fig.~\ref{fig:wavenumbers}.  The dashed line corresponds to the
    power law $\xi \propto \epsilon^{-1/2}$.}
  \label{fig:corr}
\end{figure}

We have also computed the correlation length $\xi(\epsilon)$ of the
patterns as a function of $\epsilon$.  The correlation length, defined
here as the inverse of the standard deviation of the probability
density function $P(k)$, is a measure of the average length scale of
correlated regions in the pattern.  In Fig.~\ref{fig:corr}, we show
$\xi(\epsilon)$ calculated for both unquenched patterns (denoted by
the circle symbols in Fig.~\ref{fig:corr}) and for patterns observed
when the mean flow is quenched (cross symbols).  For comparison
purposes, we have also included the correlation lengths calculated
from a previous experiment \cite{MBCA,MBCA2} performed in a
cylindrical cell with $\Gamma=78$ and $\sigma=0.95$ (diamond symbols).
We see that the correlation lengths for the states when mean flow is
quenched are, on the average, about twice as large as those for spiral
defect chaos at all values of $\epsilon$.  In addition, the
correlation lengths for the unquenched patterns can be fitted with the
power law $\xi \propto \epsilon^{-1/2}$, as has been suggested by past
experiments \cite{Hu2,MBCA2}, and which is predicted by dimensional
arguments to be valid at least near threshold.  However, the same
cannot be said for the quenched states.  In fact, the data suggests
that while an exponent of $-1/2$ might be fitted for $\epsilon
\gtapprox 0.7$, the correlation lengths appear to have saturated at
$\xi\sim \Gamma = 14$ for $\epsilon \ltapprox 0.7$.  This suggests
that finite size effects become important, and that, in order to
obtain a better estimate of the scaling relation for the patterns
observed when mean flow is quenched, we would need to use a larger
aspect ratio.  Owing to the lack of data over more decades of reduced
Rayleigh numbers, actual fittings to the data were not carried out.

\subsection{Curvature Distributions}

Finally, we quantify how much the quenching of the mean flow
straightens the rolls by looking at the distribution of the local
curvature $\chi$, defined at every point in the planform to be the
magnitude of the divergence of the unit wave vector:
\begin{equation}
  \chi = |\vec{\nabla} \dotprod \widehat{k}|.
\end{equation}
A value of $\chi=0$ corresponds to a straight roll, whereas a value of
$\chi=1$ corresponds to a roll with a radius of curvature of unity.

\begin{figure}[htb]
  \begin{center}
    \includegraphics[width=\figwidth]{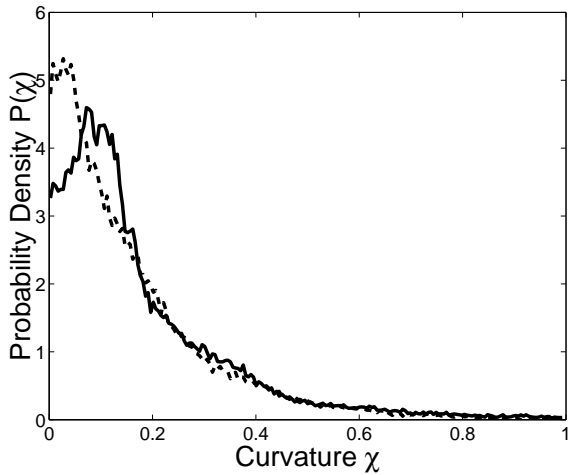}
  \end{center}
  \caption{The probability density function $P(\chi)$ of the curvature $\chi$.
    The solid line is for the spiral defect chaotic state at
    $\epsilon=1.0, \sigma=1$, and $\Gamma_x=\Gamma_y=20$ averaged over
    different random initial conditions and times $t=400$~to~$500$.
    The dashed line is for the stripes with angular bends observed at
    $10$ time units after mean flow has been quenched.}
  \label{fig:curv}
\end{figure}

We have computed the probability density function $P(\chi)$ for spiral
defect chaos observed at $\epsilon=1.0$, $\sigma=1$, and
$\Gamma_x=\Gamma_y=20$, as well as for the resulting stripes with
angular bends observed at $10$ time units after mean flow is quenched.
In Fig.~\ref{fig:curv}, we plot the two distributions.  The curvature
distribution for spiral defect chaos (solid line) peaks at a value of
$\chi \approx 0.1$, suggesting that the pattern is dominated by
spirals whose radius of curvature is $\chi^{-1} \sim 10$, consistent
visually with the pattern shown in Fig.~\ref{fig:sdc}(a).  We see that
this peak broadens to become a plateau at $0 \ltapprox \chi \ltapprox
0.1$ for the quenched state (dashed line), suggesting an increase in
the dominance of straighter rolls in the pattern.

\begin{figure}[htb]
  \begin{center}
    \includegraphics[width=\figwidth]{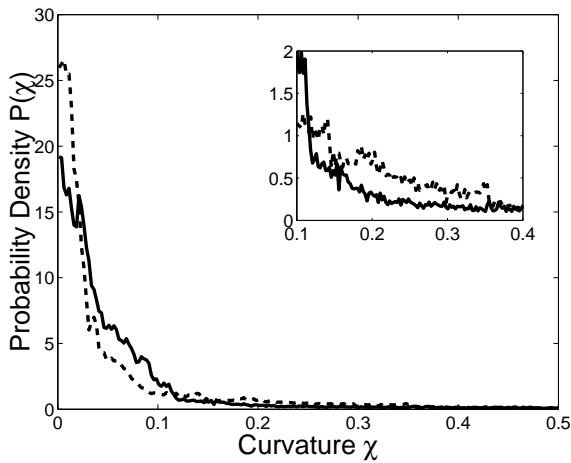}
  \end{center}
  \caption{The probability density function $P(\chi)$ of the curvature $\chi$.
    The solid line is for stripe textures at $\epsilon=0.15,
    \sigma=1$, and $\Gamma_x=\Gamma_y=20$ averaged over different
    random initial conditions and times $t=400$~to~$500$.  The dashed
    line is for the stripes with angular bends observed at $10$ time
    units after mean flow has been quenched.  Because $P(\chi) \approx
    0$ for $\chi \gtapprox 0.5$, the region $0.5 \ltapprox \chi
    \ltapprox 1$ is not plotted.  The inset is a zoom of the region
    $0.1 \le \chi \le 0.4$.}
  \label{fig:curv_stripetex}
\end{figure}

We observed similar results for the comparison of the curvature
distribution for the stripe textures.  In
Fig.~\ref{fig:curv_stripetex}, we show the comparisons for a state at
$\epsilon=0.15$.  We see that both distributions, with (solid line)
and without (dashed line) mean flow decrease approximately
monotonically and rapidly with increasing $\chi$.  Both the
comparisons for spiral defect chaos and for stripe textures suggest
that the consequence of quenching mean flow is to straighten out the
rolls.

In addition, the distribution at $\epsilon=0.15$ for the quenched case
is higher for $\chi \ltapprox 0.05$ as well as for $0.1 \ltapprox \chi
\ltapprox 0.4$ (see the inset of Fig.~\ref{fig:curv_stripetex}), and
lower otherwise.  This suggests that another consequence of quenching
mean flow is the development of angular structures that have large
curvatures.

\subsection{Mean Flow and Lateral Boundaries}

In experiments where the Rayleigh number is sufficiently high, it has
been frequently observed that convection rolls terminate
perpendicularly into the lateral walls.  We show in this section that
mean flow generated by amplitude gradients near lateral walls can be
used to explain this phenomenon, although the applicability of this
argument rests on a number of factors, among them the presence of
defects which affects the ability of the patterns to reorient
themselves.

If we call $\widehat{n}$ the outward unit vector normal to the lateral
boundary and $\widehat{k}$ the wave director of the rolls, then we can
define the wall-roll obliqueness angle as
\begin{equation}
  \label{eq:Theta}
  \Theta \equiv \arccos |\widehat{k} \dotprod \widehat{n}|.
\end{equation}
In practice, the numerical value of $\Theta$ at a particular location
along the lateral boundary is obtained by averaging
Eq.~(\ref{eq:Theta}) over a length $r=0.5$ to $r=1.5$, where $r$ is
the perpendicular distance away from that location along the lateral
boundary.  The value $\Theta=\pi/2$ corresponds to rolls terminating
perpendicularly into the walls.  The common occurrence of this value
remains a phenomenological observation, without much theoretical
understanding, although it has been found \cite{Cross_boundary} that
$\Theta$ is not fixed by the physical boundary conditions,
Eqs.~(\ref{eq:u_bcs})-(\ref{eq:T_lateral_bcs}).

\begin{figure}[htb]
  \begin{center}
    \includegraphics[width=\figwidth]{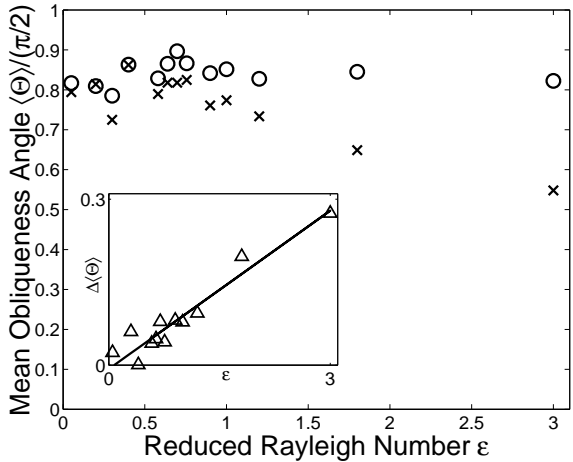}
  \end{center}
  \caption{The mean  obliqueness angle $\langle\Theta\rangle$
    as a function of the reduced Rayleigh number $\epsilon$ for states
    with mean flow (circle symbols) and with mean flow quenched (cross
    symbols).  The inset shows the difference between the two sets of
    data, $\Delta\langle\Theta\rangle$, as a function of $\epsilon$.}
  \label{fig:theta_epsilon}
\end{figure}

However, when rolls do not terminate perpendicularly at a lateral
boundary, variations in the amplitude of the convection rolls as it
decays near the lateral boundaries results in the generation of a mean
flow.  (Recall from Eq.~(\ref{eq:u_D}) that a mean flow is generated
by inhomogeneities in the wavenumbers and amplitudes of the convection
rolls.)  The normal component (with respect to the lateral boundary)
of this mean flow is cancelled by the flow generated from from slow
pressure gradients, resulting in the mean flow being parallel to the
lateral boundary.  It then tends to push the rolls back to a
perpendicular orientation.  The actual calculations are worked out in
Appendix~\ref{se:restor}.  The importance of this mean flow in
ensuring that the rolls terminate perpendicularly is indicated by
observing that, in the absence of mean flow, oblique rolls are more
prevalent.  In Fig.~\ref{fig:theta_epsilon}, we plot the wall-roll
obliqueness angle averaged over the lateral boundaries for patterns
observed at $t=500$ at various reduced Rayleigh number $\epsilon$,
Prandtl number $\sigma=1$ and in a rectangular cell of aspect ratio
$\Gamma_x=\Gamma_y=20$, with mean flow and with the mean flow
quenched.  We see that, with mean flow, the rolls are close to
perpendicular, $\Theta \approx \pi/2$.  However, when mean flow is
quenched, the rolls are more oblique, $\Theta \ltapprox \pi/2$.  In
fact, the difference in the mean obliqueness between the states with
mean flow and with mean flow quenched, $\Delta\langle\Theta\rangle$,
increases approximately linearly with $\epsilon$,
\begin{equation}
  \Delta \langle\Theta\rangle = (0.16 \pm 0.01) \epsilon,
\end{equation}
as the inset of Fig.~\ref{fig:theta_epsilon} depicts.

\begin{figure}[htb]
  \begin{center}
    \includegraphics[width=\figwidth]{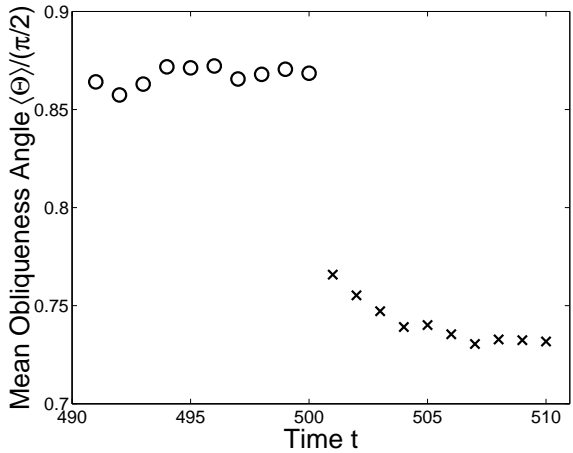}
  \end{center}
  \caption{The change in the mean obliqueness angle
    $\langle\Theta\rangle$ as a function of time, averaged over
    different random initial conditions.  The parameters here are
    $\epsilon=1.0$, $\sigma=1$, and $\Gamma_x=\Gamma_y=20$.  Mean flow
    quenching takes place at time $t=t_q=500$, so that, for $490\le
    t\le 500$, the mean wall-roll obliqueness angle is for a pattern
    whose bulk dynamics exhibits spiral defect chaos, whereas for $500
    \le t\le 510$, the bulk dynamics is made up of stripes with
    angular bends.}
  \label{fig:theta_evolve}
\end{figure}

When mean flow is quenched, the reorientation of the rolls away from
$\Theta=\pi/2$ is almost instantaneous.  We illustrate this result in
Fig.~\ref{fig:theta_evolve} for one particular reduced Rayleigh
number, $\epsilon=1.0$.  In this case, the mean flow quenching takes
place at time $t=t_q=500$.  We see that, at time $t=500$, the mean
wall-roll obliqueness angle moves away from $\Theta=\pi/2$ in a time
scale of $\mathcal{O}(1)$.

\begin{figure}[htb]
  \begin{center}
    \includegraphics[width=\figwidth]{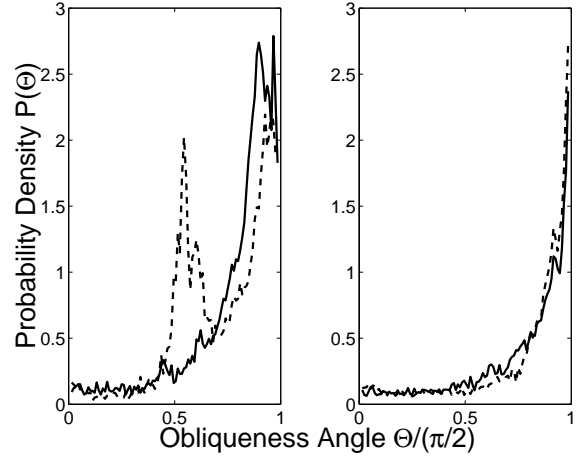}
  \end{center}
  \caption{\emph{(a)} Distribution of angles that rolls terminate at a lateral
    boundary in a rectangular cell of aspect ratio
    $\Gamma_x=\Gamma_y=20$.  The solid line shows the distribution for
    the spiral defect chaotic state averaged over different initial
    conditions at $\epsilon=1.0$ and $\sigma=1$.  The dashed line
    shows the distribution for the state with mean flow quenched.
    \emph{(b)} The solid line shows the distribution for spiral defect
    chaos observed in a cylindrical cell of aspect ratio $\Gamma=30$
    at $\epsilon=1.0$ and $\sigma=1$.  The dashed line shows the
    distribution for this state but with mean flow quenched.}
  \label{fig:theta}
\end{figure}

The above argument that mean flow restores the rolls to a
perpendicular orientation may not always be applicable.  For
example, when we performed simulations in a cylindrical cell of aspect
ratio $\Gamma=30$, we find that, at $\epsilon=1.0$, the mean
obliqueness angle $\langle\Theta\rangle$ still remains close to
$\pi/2$ when mean flow is quenched.  This can be seen more clearly in
Fig.~\ref{fig:theta}, where we show the probability density
$P(\Theta)$ of obliqueness angles along the lateral boundaries for
states observed in a rectangular cell of aspect ratio
$\Gamma_x=\Gamma_y=20$ and cylindrical cell of aspect ratio
$\Gamma=30$.  We see that in a cylindrical cell with mean flow
quenched, the peak at $\Theta\approx\pi/2$ is still observed after the
mean flow has been quenched.  One possible explanation might be that,
in a cylindrical cell, there are more defects existing near
the lateral boundaries and that these defects then pin the rolls,
preventing them from reorienting  away from $\Theta=\pi/2$
when the mean flow is quenched.

\begin{figure}[htb]
  \begin{center}
    \includegraphics[width=\figwidth]{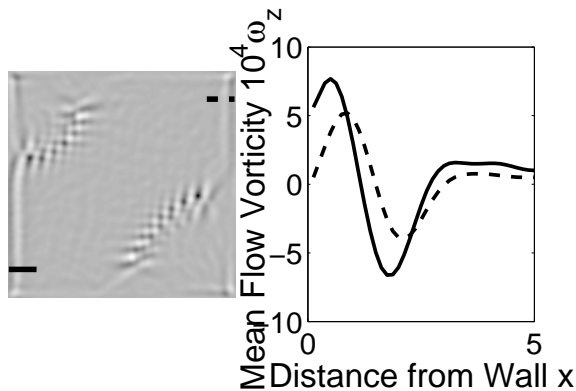}
  \end{center}
  \caption{\emph{(a)} The mean flow vorticity $\omega_z(x,y)$
    corresponding to the stripe texture of Fig.~\ref{fig:stripetex}(a)
    obtained using Eq.~(\ref{eq:zeta}).  Light regions correspond to
    positive vorticity, dark regions to negative vorticity.  The
    important feature in this vorticity map is the presence of
    ``jet''-like structures along the bottom half of the left wall and
    the top half of the right wall.  \emph{(b)} The vorticity
    $\omega_z(x)$ is plotted along the solid and the dashed
    horizontal lines shown in (a).  The shape of $\omega_z(x)$ is to
    be compared with Fig.~(\ref{fig:ftx}) in Appendix~\ref{se:restor},
    where a positive and a negative vorticity patch sets up a
    restoring mean flow.}
  \label{fig:stripetex_vp2}
\end{figure}

Another instance where the above argument does not apply is at low
Rayleigh numbers.  From Eq.~(\ref{eq:U_restor}) in
Appendix~\ref{se:restor}, the magnitude of the mean flow $|\vec{U}|
\propto \epsilon^{1/2}$ so that at low Rayleigh numbers, the mean flow
may not be strong enough to reorient the rolls perpendicularly.  This
is evident in Fig.~\ref{fig:stripetex}(a), where, at the reduced
Rayleigh number $\epsilon=0.15$, rolls are seen to terminate with an
acute angle at the lateral walls.  In this case, the presence of a
restoring mean flow can also be visualized.  The mean flow vorticity
plot corresponding to this pattern, shown in
Fig~\ref{fig:stripetex_vp2}(a), shows the presence of strong vorticity
along the bottom half of the left wall and the top half of the right
wall.  There, the restoring mean flow vorticity takes the form of long
and narrow circulating ``jets'' that are about one roll size wide and
several roll sizes long.  In Fig.~\ref{fig:stripetex_vp2}(b), the
vorticity is plotted as a function of distance away from the lateral
wall along the solid and dashed horizontal lines shown in
Fig.~\ref{fig:stripetex_vp2}(a).  The existence of a positive
vorticity patch close to the wall and a negative patch further away
from the wall, which together indicates the presence of a restoring
mean flow, agree qualitatively with the theoretical results of
Fig.~\ref{fig:ftx} in Appendix~\ref{se:restor}.  When the Rayleigh
number of the state in Fig.~\ref{fig:stripetex} is increased from
$\epsilon=0.15$ to $\epsilon=1.0$, the mean flow becomes strong enough
to reorient the rolls to become perpendicular to the lateral walls,
and subsequently disappears.

\section{Conclusion}
\label{se:con}

In this article, we have described a novel procedure to construct a
modified velocity field that does not have any mean flow in a
convecting flow.  We have applied this procedure to show that spiral
defect chaos does not survive when the mean flow is quenched.
Instead, a pattern characterized by textures of stripes with angular
bends appears.  We have also shown that the mean wavenumbers of these
quenched patterns approach those selected by focus-type singularities,
which, in the absence of mean flow, lie at the boundary of the zig-zag
instability.

We next presented a heuristic argument on how mean flow can contribute
to rolls terminating into a lateral boundary perpendicularly.  We
provided data to show that, in the absence of mean flow, rolls begin
to deviate from a perpendicular orientation, and this obliqueness
increases with Rayleigh number.  However, the ability of this mean
flow to restore the rolls to a perpendicular orientation may be
impeded by the presence of defects that do not allow the rolls to
reorient themselves, and at low Rayleigh numbers where the restoring
mean flow is weak.

\begin{acknowledgments}

This work is supported by the Engineering Research Program of the
Office of Basic Energy Sciences at the Department of Energy, Grants
DE-FG03-98ER14891 and DE-FG02-98ER14892.  We acknowledge the Caltech
Center for Advanced Computing Research and the North Carolina
Supercomputing Center.  We also thank Paul Fischer for helpful
discussions.

\end{acknowledgments}

\appendix
\section{Derivation of Quenching of Mean Flow}
\label{se:fudge}

In this appendix, we derive the functional form of the forcing term
$\vec{\Phi}$ that is to be added to the fluid equation,
Eq.~(\ref{eq:NS}), to make the resulting fluid dynamics have
zero mean flow.

As mentioned in Section~\ref{se:meanflow}, the mean flow comprises a
local component generated by the Reynolds stress $1/(2\pi)
\int_0^{2\pi} d\phi \vec{u} \dotprod \vec{\nabla}_\perp \vec{u}_\perp$
and a global component driven by slow a horizontal pressure gradient
that is present in order to guarantee the incompressibility condition,
Eq.~(\ref{eq:incom}).  Thus, if the Reynolds stress is subtracted from
the dynamics at all times, then mean flow will not be generated.  We
thus suggest that
\begin{equation}
  \Phi(x,y,t) = \frac{1}{2\pi} \int_0^{2\pi} d\phi\, \rho \int_{-1/2}^{1/2} dz
\vec{u} \dotprod \vec{\nabla}_\perp \vec{u}_\perp,
\end{equation}
where the operator $\rho \int_{-1/2}^{1/2} dz$ serves as an average
over the depth of the cell.  This $\vec{\Phi}$ can then be subtracted
from the fluid equation, Eq.~(\ref{eq:NS}), resulting in
Eq.~(\ref{eq:NS_imaginary}).

We now need to evaluate the value of the constant $\rho$.  To do this,
we rewrite the equation for the slow distortions, Eq.~(\ref{eq:u_D}),
as
\begin{equation}
  \label{eq:u_D_A2}
  \sigma \partial_{zz} \vec{u}_D = \vec{\nabla}_\perp p_s +
  \frac{1}{2\pi} \int_0^{2\pi} d\phi\, \vec{u} \dotprod \vec{\nabla}
  \vec{u}_\perp - \vec{\Phi}.
\end{equation}
Following Ref.~\cite{CN}, the Reynolds stress term near threshold
takes the form
\begin{equation}
  \frac{1}{2\pi} \int_0^{2\pi} d\phi\, \vec{u} \dotprod \vec{\nabla}
  \vec{u}_\perp \equiv I(k,z) \vec{R}(x,y).
\end{equation}
where
\begin{equation}
  \vec{R}(x,y) \equiv \vec{k} \vec{\nabla}_\perp \dotprod (\vec{k} A^2),
\end{equation}
and
\begin{equation}
  I(k,z) \equiv  w_0(k,z) \partial_z \frac{\partial \phi_0(k,z)}{\partial
  k_c^2} - \frac{\partial w_0(k,z)}{\partial k_c^2}\partial_z \phi_0(k,z),
\end{equation}
with $w_0(k,z)$ and $\phi(k,z)$ the vertical profiles of the vertical
velocity and the potential of the horizontal velocities, respectively.
For systems satisfying the rigid boundary condition,
Eq.~(\ref{eq:u_bcs}), these functions are the familiar Chandrasekhar
functions \cite{Chandra}.

We can then rewrite Eq.~(\ref{eq:u_D_A2}) as
\begin{eqnarray}
  \label{eq:u_D3}
  \sigma \partial_{zz} \vec{u}_D &=& \vec{\nabla}_\perp p_s + I(k,z)
  \vec{R}(x,y) \nonumber \\
    & & - \rho  \int_{-1/2}^{1/2} dz I(k,z) \vec{R}(x,y).
\end{eqnarray}
Integrating Eq.~(\ref{eq:u_D3}) with respect to $z$ twice, and making use of
the boundary condition, Eq.~(\ref{eq:u_bcs}),
\begin{eqnarray}
  \sigma \vec{u}_D &=& p(z) \vec{\nabla}_\perp p_s + J(k,z)
  \vec{R}(x,y) \nonumber \\
  & & - p(z) \rho \int_{-1/2}^{1/2} dz I(k,z) \vec{R}(x,y),
\end{eqnarray}
with
\begin{equation}
  p(z) \equiv \frac{1}{2}\left(z^2-\frac{1}{4}\right)
\end{equation}
the Poiseuille profile, and $J(k,z)$ the double integral of $I(k,z)$
with respect to $z$.  Employing the incompressibility condition,
Eq.~(\ref{eq:u_D_incom}), we then arrive at the equality
\begin{equation}
  \rho = \frac{12 \int_{-1/2}^{1/2} dz J(k,z)}{\int_{-1/2}^{1/2} dz I(k,z)}.
\end{equation}
Evaluating these integrals numerically yields $\rho\approx 1.5$ for
the rigid boundary condition, Eq.~(\ref{eq:u_bcs}).  Moreover, $\rho$
is relatively independent of the wavenumber $k$, varying from
$\rho=1.4886$ at $k=2.8$ to $\rho=1.4887$ at $k=k_c=3.117$ and to
$\rho=1.4886$ at $k=3.4$, suggesting the validity of treating it as a
constant.

\begin{figure}[htb]
  \begin{center}
    \includegraphics[width=\figwidth]{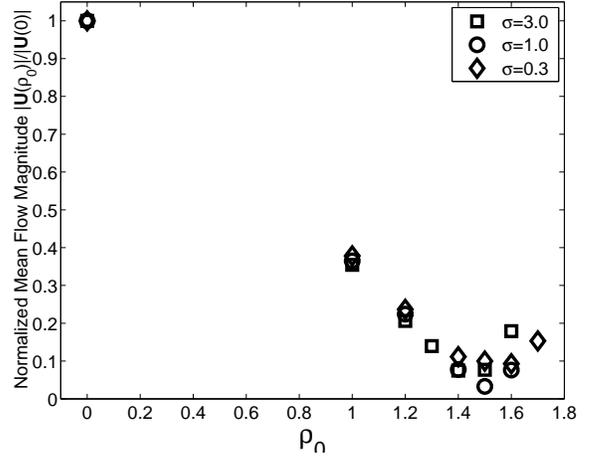}
  \end{center}
  \caption{The maximum mean flow magnitude vs. various trial values
  of $\rho$.  The mean flow magnitudes are normalized by their
  values at $\rho=0$, i.e., when there is no quenching.  We see that,
  when $\rho \approx 1.5$, the mean flow goes to zero, for all three
  Prandtl numbers.}
  \label{fig:fudge}
\end{figure}

Finally, to numerically confirm this result, we carry out the quenching of the
mean flow, as described in Eq.~(\ref{eq:NS_imaginary}), for a range of values
for $\rho$, at $\epsilon=1.0$ in a rectangular cell of $\Gamma_x=\Gamma_y=20$.
At $10$ time units after effecting the quenching, we then measure the maximum
magnitude of the mean flow as a function of $\rho$.  We plot our results in
Fig.~\ref{fig:fudge}, where we show the maximum mean flow magnitude (normalized
by the maximum mean flow magnitude observed without quenching) vs. $\rho$ for
data from three different Prandtl numbers.  We see that, when $\rho \approx
1.5$, the normalized maximum mean flow magnitude is indeed zero.

\section{Restoring Mean Flow Near a Lateral Boundary}
\label{se:restor}

\begin{figure}[htb]
  \begin{center}
    \includegraphics[width=\figwidth]{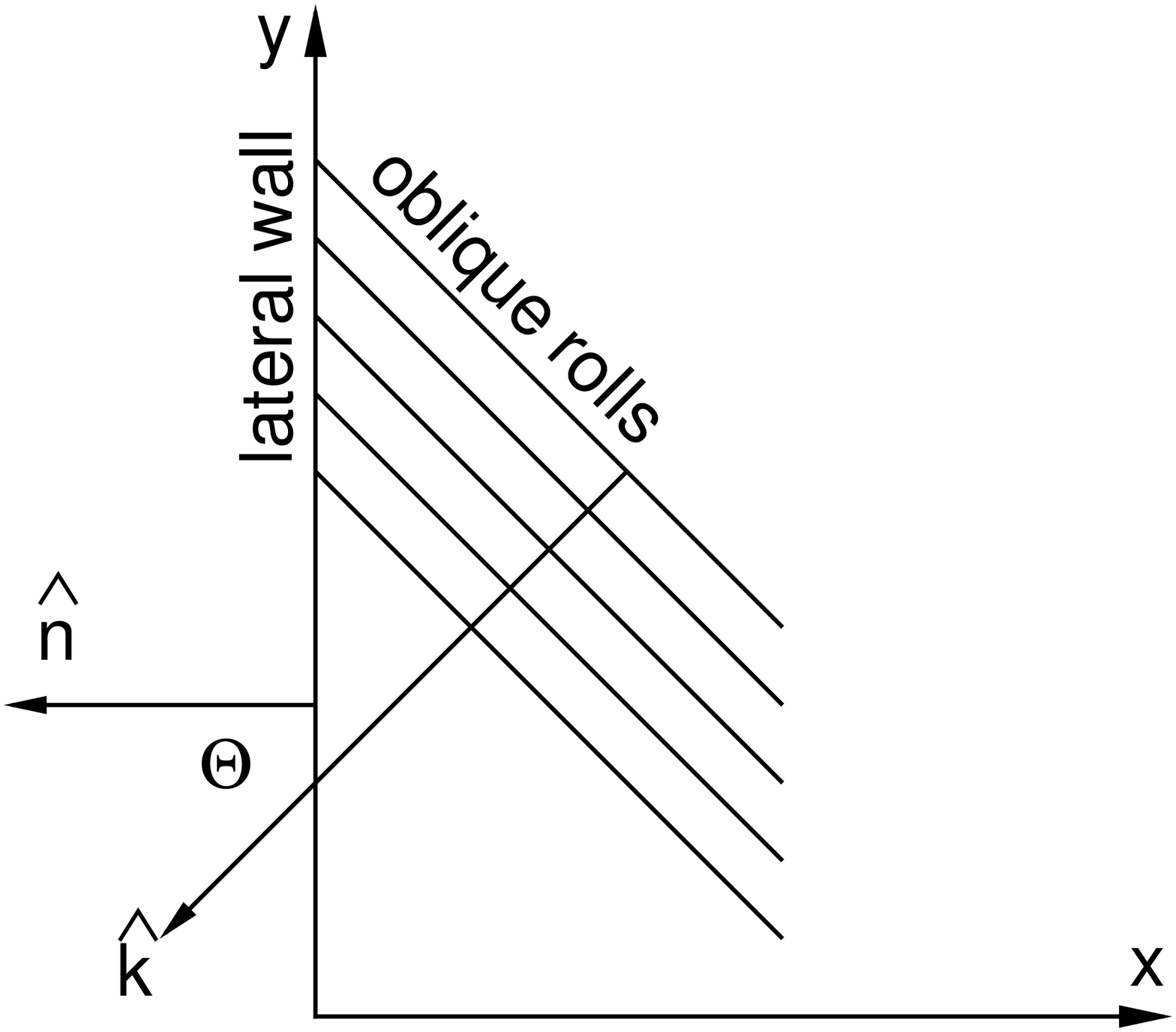}
  \end{center}
  \caption{Straight and parallel convection rolls with wave
    director $\widehat{k}$ terminating at a lateral boundary with
    outward normal $\widehat{n}$ at an angle of obliqueness $\Theta$.
    Note that, by our definition, $0 < \Theta <\pi/2$.  The
    perpendicular distance away from the lateral wall is $x$.}
  \label{fig:sketch}
\end{figure}

In this appendix, we show that a set of straight and parallel rolls
that are oriented obliquely at an angle $\Theta$ to a lateral boundary
sets up a mean flow that tends to restore the rolls back to being
perpendicular to the lateral boundary.  The various quantities used
here are defined in the sketch in Fig.~\ref{fig:sketch}.

We will make the assumption that the wave vectors of the rolls are
constant near the lateral boundary,
\begin{equation}
  \label{eq:k}
  \vec{k} = (-k\cos\Theta,-k\sin\Theta),
\end{equation}
and that the convection amplitude within a correlation length $\xi$ of
a lateral boundary is suppressed \cite{the_review,Wesfreid}:
\begin{equation}
  \label{eq:A2}
  A(x,y) = A_0 \tanh\left( \frac{x}{\xi \cos\Theta} \right).
\end{equation}
The quantity $A_0$ is the amplitude in the bulk.  The correlation
length $\xi=\sqrt{2} \epsilon^{-1/2}\xi_0$ with $\xi_0=0.385$.  The
variable $x$ is the perpendicular distance away from the lateral
boundary.

\begin{figure}[htb]
  \begin{center}
    \includegraphics[width=\figwidth]{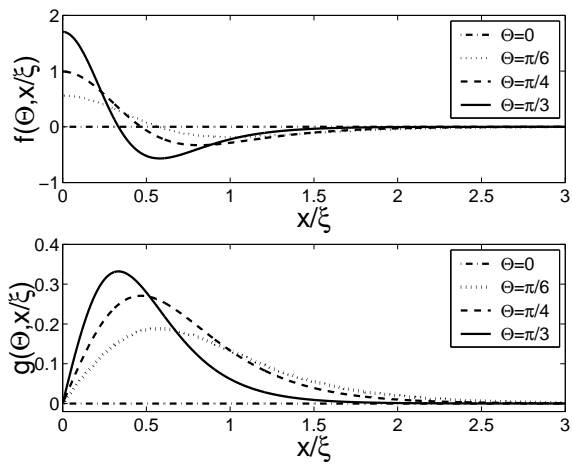}
  \end{center}
  \caption{\emph{(a)} The function $f(\Theta,x)$ defined in
    Eq.~(\ref{eq:ftx}) which is the normalized vertical component of
    the mean flow vorticity.  It is plotted here for several values of
    $\Theta$ and vs. $x/\xi$.  \emph{(b)} The function $f(\Theta,x)$
    defined in Eq.~(\ref{eq:gtx}) which is the normalized mean flow
    magnitude.  It is plotted here for several values of $\Theta$ and
    vs. $x/\xi$.}
  \label{fig:ftx}
\end{figure}

Then, from the Cross-Newell equation \cite{CN}, the amplitude
gradients near the lateral wall will result in a non-zero mean flow
vorticity, $\omega$, given by
\begin{equation}
  \label{eq:CNzeta}
  \omega = \gamma \widehat{z} \dotprod \vec{\nabla}_\perp \times [
  \vec{k} \vec{\nabla}_\perp \dotprod (\vec{k} A^2)],
\end{equation}
where $\gamma$ is a constant that is inversely proportional to the
Prandtl number $\sigma$.  (If we relax the assumption that the
wavenumbers of the rolls are constant, then the compression and
dilation of the rolls as well as inhomogeneities in their curvatures
will also contribute to the mean flow.)  Substituting
Eqs.~(\ref{eq:k})~and~(\ref{eq:A2}) into Eq.~(\ref{eq:CNzeta}) then
gives
\begin{equation}
  \omega(x) = 2 \gamma A_0^2 k^2 \xi^{-2} f(\Theta,x)
\end{equation}
where the normalized mean flow vorticity,
\begin{eqnarray}
  \label{eq:ftx}
  f(\Theta,x)&=&\tan(\Theta) \sech^2\left( \frac{x}{\xi \cos\Theta} \right) \\
  & & \times  \left[1  - 3\tanh^2\left( \frac{x}{\xi \cos\Theta}
  \right) \right], \nonumber
\end{eqnarray}
is plotted in Fig.~\ref{fig:ftx}(a) for several representative values
of $\Theta$.  We see that $\omega$ is positive for $x/\xi \ltapprox
1$, and negative otherwise.  The currents from this vorticity pair
will then drive the rolls back to a perpendicular orientation.

The mean flow generated by this vorticity can also be easily computed.
Along the lateral wall, it is given by
\begin{equation}
  |\vec{U}| = \gamma |\vec{k}_y| \vec{\nabla}_\perp \cdot (\vec{k} A^2).
\end{equation}
(The component of the mean flow normal to the lateral wall is
cancelled by the flow coming from the slow pressure gradient.)  Using
Eqs.~(\ref{eq:k})~and~(\ref{eq:A2}), we arrive at
\begin{equation}
  \label{eq:U_restor}
  |\vec{U}| = 2\gamma A_0^2 k^2 \xi^{-1} g(\Theta,x)
\end{equation}
where the normalized restoring mean flow magnitude in the direction of
the lateral wall
\begin{equation}
  \label{eq:gtx}
 g(\Theta,x)= \sin(\Theta) \sech^2\left( \frac{x}{\xi
 \cos\Theta}\right)\tanh\left(\frac{x}{\xi \cos\Theta}\right).
\end{equation}
is plotted in Fig.~\ref{fig:ftx}(b) for several representative values
of $\Theta$.

\begin{figure}[htb]
  \begin{center}
    \includegraphics[width=\figwidth]{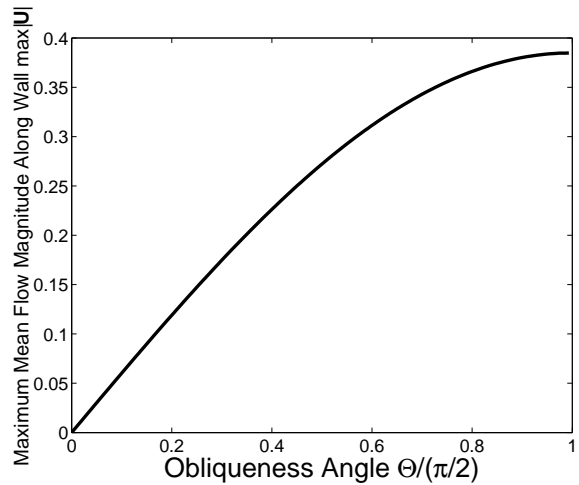}
  \end{center}
  \caption{The maximum magnitude of the  mean flow
    $\vec{U}$ as a function of the wall-roll obliqueness angle
    $\Theta$.  It increases monotonically from zero at $\Theta=0$
    (rolls parallel to the wall).}
  \label{fig:max_U_R}
\end{figure}

Finally, we plot the quantity $\max |\vec{U}|$ as a function of
$\Theta$ in Fig.~\ref{fig:max_U_R}.  We see that the restoring mean
flow magnitude grows monotonically from zero at $\Theta=0$
(corresponding to sets of rolls parallel to the lateral wall) to
attain its largest value at $\Theta \rightarrow \pi/2$ (corresponding
to sets of rolls perpendicular to the wall).  Our analysis actually
breaks down for $|\Theta-\pi/2| \ltapprox \epsilon^{1/4}$ because
modifications at the next order in Eq.~(\ref{eq:A2}) become important
\cite{Cross_boundary}.

\end{document}